\documentclass[11pt,twoside]{article}
\usepackage{graphicx,epsfig,natbib,epstopdf}
\usepackage{CS18}
\begin{document}
%
% NTA: Ignore the setcounter comment here
%\setcounter{page}{7}
%
% NTA: Update your full title here
%
\title{Magnetic Interaction of a Super-CME with the Earth's
Magnetosphere: Scenario for Young Earth}
%
% NTA: enter your author name, affiliation/address information here
%
\author{V. S. Airapetian$^{1}$, A. Glocer$^{2}$, W. Danchi$^{3}$}
\affil{$^1$NASA/Goddard Space Flight Center, Greenbelt, MD 20771}
%\affil{$^2$National Optical Astronomy Observatories, P.O. Box 26732, Tucson, Arizona, USA 85726}
%\affil{$^3$Institut d'Astrophysique de Paris, 98bis bd Arago, 75014 Paris, France}
%
\begin{abstract}
Solar eruptions, known as Coronal Mass Ejections (CMEs), are frequently
observed on our Sun. Recent Kepler observations of superflares on G-type stars have implied that so called
super-CMEs, possessing kinetic energies 10 times of the most powerful CME
event ever observed on the Sun, could be produced with a frequency of 1 event
per 800-2000 yr on solar-like slowly rotating stars. We have performed a 3D time-dependent global
magnetohydrodynamic simulation of the magnetic interaction of such a CME
cloud with the Earth's magnetosphere. We calculated the global structure of
the perturbed magnetosphere and derive the latitude of the open-closed magnetic field boundary.
We also estimated energy fluxes penetrating the Earth's
ionosphere and discuss the consequences of energetic particle fluxes on biological systems on early Earth.
\end{abstract}
%
%
%
%
% NTA: Here is where the body of your text goes.
%
\section{Introduction}

%A piece of advice: delete the fake Latin before inserting your own text and then compiling the LaTeX/PDFLaTeX.
Coronal mass ejections (CMEs) represent large scale and energetic solar eruptions that serve as the primary drivers of geomagnetic storms and other space weather disturbances on Earth. Observations of CME dynamics with SOHO and STEREO coronagraphs suggest that they are formed in the solar corona as a result of a large-scale magnetic field restructuring. As these magnetic structures propagate out into inteplanetary space with speeds as high as 3000 km/s and hit the Earth's magnetosphere, they produce extremely intensive fluxes of solar proton events (SEP) and magnetic storms (Gopalswamy 2011). The Carrington magnetic storm occurred on September 1-2, 1859 represents one of the most powerful storms in recorded history. According to historical records, this storm was caused by a CME event with the kinetic energy of the order of 2$\times$10$^{33}$ erg. Its impact on Earth's magnetosphere had initiated extensive auroral events visible at low latitudes as low as 19 degrees, fires and electrical shocks due to strong induced electric currents (Tsurutani \& Lakhina 2014). Much larger SEP event occurred in AD 775 as implied by tree ring data (Miyake et al. 2012) can also be attributed to a CME event of a series of events with the energy of about 2$\times$10$^{33}$ erg if we assume the opening angle of 24 deg (Thomas et al. 2013). However, the statistical distribution of a CME width increases with their energies and reaches up to 100 deg. If we assume a representative opening angle of 80 deg, the estimate of the energy of AD 775 CME event increases by one order of magnitude. The frequency of occurrence of such energetic events in the Sun is estimated as 2$\times$10$^{-4}$ - 1.8$\times$10$^{-3}$, which is relatively small number. What effects can be expected for our civilization and life on Earth if such an event hits the Earth? How frequently and how detrimental were the effects of CMEs from young Sun to Hadean Earth? CMEs can not be directly observed from other solar-like stars except for possible type III and type IV  bursts at decameter wavelength introduced by accelerated electrons as a CME propagates out from the solar/stellar corona (Boiko et al. 2012; Konovalenko et al. 2012). However, The frequency of CMEs from young Sun and other active stars can be estimated from their association with solar/stellar flares. Recent SOHO/LASO and STEREO observations of energetic and fast ($\geq$ 500 km/s) CMEs show strong association with powerful solar flares (Yashiro \& Gopalswamy 2009; Aarnio et al. 2011; Tsurutani \& Lakhina 2014). This empirical correlation provides a direct way to characterize CME frequencies of occurrence from statistics of solar and stellar flares. Recently, Kepler observations detected superflares on K-G type main-sequence stars with energies within (1-100) $\times$ 10$^{34}$ ergs (Shibayama et al. 2013). In the meantime, solar observations of energetic solar events suggest an association of CMEs with energetic solar flares and Solar Energetic Particle events (SEP) that produce energetic protons with energies $>$700 MeV (Aarnio et al. 2011; Thakur et al. 2014). This indicates that magnetically active young Sun (0.5 Ga) could produce frequent and powerful CMEs with energies $\geq$~10$^{34}$ ergs we will refer to as super-CMEs.

In Section 2 we derive the occurrence frequency of a "regular" super-CME event on young Sun. In Section 3 we discuss the results of our 3D MHD model of interaction of a "regular" CME events with the magnetosphere of a young Earth. Section 4 discusses the results of our study and its implications for life conditions on early Earth and Earth twins.

\section{Frequency of Occurrence of Super-CMEs from Young Sun}

Our Sun is a magnetically active star that exhibits CMEs with kinetic energies in the range between 3$\times$10$^{31}$ erg to 10$^{33}$ ergs. Because the physical nature of a CME's origin is not well understood, we need to rely mostly on statistics for their properties. With the advent of SOHO/LASCO and later STEREO missions, our understanding of statistical relationship between physical parameters of CMEs and their frequencies of occurrence provided quantitative predictive capabilities. Energetic flares are found be occurred around the times of initiation of extremely powerful CMEs (Aarnio et al. 2011; Nitta et al. 2014). Recent studies suggests that fast ($>$ 900 km/s) and wide ($\theta >$ 100 deg) CMEs show association with solar flares. We will use this relationship in order to estimate the frequency of occurrence of our Sun 4 billion years ago, when it represented a rapidly rotated magnetically active star. According to the empirical trends of magnetic flux generation with age, the unsigned average surface magnetic field, $<B_{av}>$, on young Sun should be at least 10 times greater than that observed in the current Sun (Vidotto et al. 2014). However, this value does not account for the filling factors, and, therefore, the magnetic field strengths in active regions.  We can derive this value from the empirical nearly linear relationship between the magnetic flux, $\Phi$, observed in the Sun and stars and X-ray luminosity, $L_X$, produced in their coronae (Pevtsov et al. 2003)
\begin{center}
$L_X \propto {\Phi}^{\alpha}$,
\end{center}
where the power law index $\alpha=\rm 1.13 \pm 0.05$. This relationship holds for solar/stellar data spanning 12 orders of magnitude. By assuming a typical value of the X-ray-to-bolometric luminosity ratio, $L_x/L_{bol} \approx -3.5$, and applying this relationship to the young Sun, we find that the magnetic flux in active regions should be about 300 times larger compared to the present value. This suggests that stronger magnetic flux generated more energetic and frequent flare and CME events. Indeed, the analysis of recent Kepler observations of 279 G-type main-sequence stars reveals flare activity with energies between 10$^{34}$ - 10$^{36}$ ergs referred to as superflares. This is as high as 4 orders of magnitudes greater than that observed on the Sun. The statistics of Kepler data suggests that the frequency of occurrence of the superflares observed on G-types dwarfs follow the power-law distribution with the spectral index in the range between 2.0 - 2.2, which is similar to those observed on dMe stars and the Sun (Gershberg 2005; Shibayama et al. 2013). The occurrence rate of superflares with energies, $> 5\times10^{34}$ erg, for young solar-like stars is higher than 0.1 events per day. If we extrapolate this value for a flare energy of $E > E_0$=10$^{33}$ erg, we obtain the frequency of occurrence that is 2500 times greater, or 250 flare events per day! If every superflare with $E > E_0$ is associated with a CME event, then we obtain that the early Sun was generating 250 super-CMEs per day. Energy balance in observed flares and associated CME events suggests that the kinetic energy of CMEs carries at least 10-50 times more energy that the total energy in the associated flare event. Measurements of speeds and magnetic field strengths of propagating magnetic clouds in the interplanetary space (so called Interplanetary CMEs or ICMEs) show that the magnetic energy density in an ICME event can be comparable or larger than its kinetic energy). Therefore, we can assume that a super-CMEs possesses a total (kinetic+magnetic) energy of 5 $\times$ 10$^{34}$ ergs. This energy is a factor of 2-3 greater than that suggested for the famous Carrington-type CME event (Tsurutani \& Lakhina 2013).

Next, we assume that the CMEs produce a random orientation of magnetic field in the ejecting flux rope with respect to the Earth's dipole magnetic field. We can use these estimate to determine the rate of impacts of such super-CMEs with the early Earth's magnetosphere. To determine that, we will apply the statistical correlations established for solar CMEs between the observed widths of a flare-associated solar CMEs and its mass and speed or its total kinetic energy (Belov et al. 2014). This correlations suggests that more energetic CMEs have larger widths. Next, the kinetic energy of a CME is scaled with associated flare X-ray fluence (Yashiro \& Gopalswamy 2009). This correlations suggest the width of 80$^{\circ}$ for CMEs associated with strongest, X class solar flares. For stronger events like the one we consider for a super-CMEs, the width should be proportionally higher. But if we conservatively assume that a SCME propagates within a cone of 80$^{\circ}$, then its probability to hit the Earth is about 5\%. Also, assuming a configuration for a "perfect storm" when the incoming magnetic field is sheared with respect to the Earth's magnetic field, we obtain that a conservative estimate for frequency of impacts of a SCME with Earth as greater than 1 events per day! Thus, we conclude that early Earth was constantly exposed to superstorms that were larger than the largest superstorm observed in September 1-2, 1959! What signatures would such a SCME produce in the Earth's magnetosphere and ionosphere?

\section{3D MHD model of SCME interaction with Young Earth}

\subsection{Model Setup}

In this paper, we utilized a single-fluid, time dependent fully non-linear 3D magnetohydrodynamic (MHD) code BATSRUS for fullly ionized plasma developed by the University of Michigan which is coupled to Rice Convection Model (RCM, De Zeeuw et al. 2004) to model a propagation and interaction of a model SCME with a magnetosphere and ionosphere of a young Earth-like planet. The MHD part of the code calculates the dynamic response of the large-scale magnetospheric plasma to varying solar wind conditions in a self consistent manner by using the block-adaptive wind Roe-type upwind scheme global MHD code (Powell et al. 1999). The dynamics of the magnetosphere is described in a Cartesian geometry by using a resistive MHD equations. The electromagnetic coupling of the magnetosphere to a conducting ionosphere is handled in a standard way (Ridley et al. 2002). Specifically, the magnetospheric currents near the inner boundary of the MHD simulation are mapped to the ionosphere where. A potential solver is then used which combines these currents with a conductance map of the ionosphere (including solar and auroral contributions) to produce the electric potential in the ionosphere. That potential is then used to set the electric field and corresponding drift at the magnetospheric simulations inner boundary.

The MHD approximation not provide an adequate description of the the inner magnetosphere because energy
dependent particle drifts and ring current evolution become important. Here we use the Rice Convection
Model, embedded in the MHD simulation, to model this important region (De Zeeuw et al. 2004). The RCM code is a kinetic plasma model that couples plasma motions in the inner magnetosphere and calculates the energy dependent particle drifts and ring current evolution in the inner magnetosphere. The ring current carries the most of the energy density during magnetic storms and is essential to modeling strong storms. This coupling is crucial for description of solar wind effects on a magnetosphere, because the ionosphere provides closure of magnetospheric currents, which is needed for realistic description of magnetospheric convection and associated electric fields. Thus, we apply a dedicated inner magnetospheric model that is fully coupled to the MHD code for the treatment of the inner magnetosphere. We simulate the magnetospheric cavity (outer and inner magnetosphere) in a computational box defined by the following dimensions
$-224~R_E < x < 224 R_E, -128~R_E < y < 128 R_E$, $-128~R_E < z < 128 R_E$, where $R_E$ is the radius of the Earth placed at the center of the computational box. The dipole tilt is neglected in this problem. The simulations were carried out using a block adaptive high resolution grid with the minimum cell size of 1/16 $R_E$.

The inner boundary is set at 1.25 $R_E$ with a density of 100 cm$^{-3}$. The velocity at the inner body is set to the $\vec{E}\times\vec{B}$ velocity, where $\vec{E}$ is determined from the ionospheric potential and B is the Earth's magnetic field. The pressure is set to float. The magnetic field is set in a way that the radial component is the Earth's dipole and the tangential components are allowed to float. The simulation is initialized with a dipole everywhere in the computational domain and a small density, zero velocity, and a finite pressure. The solar wind conditions are set at the upstream boundary and some period of local time stepping is used to get an initial steady state solution.

We assume the solar wind input parameters including the three components of interplanetary magnetic field, $B_x$, $B_y$ and $B_z$, the plasma density and the wind velocity, $V_x$, using the physical conditions associated with a Carrington-type event as discussed by Tsurutani et al. (2003) and Ngwira et al. (2014) (see Figure 1).

\begin{figure}
\centering
\includegraphics*[width=\linewidth]{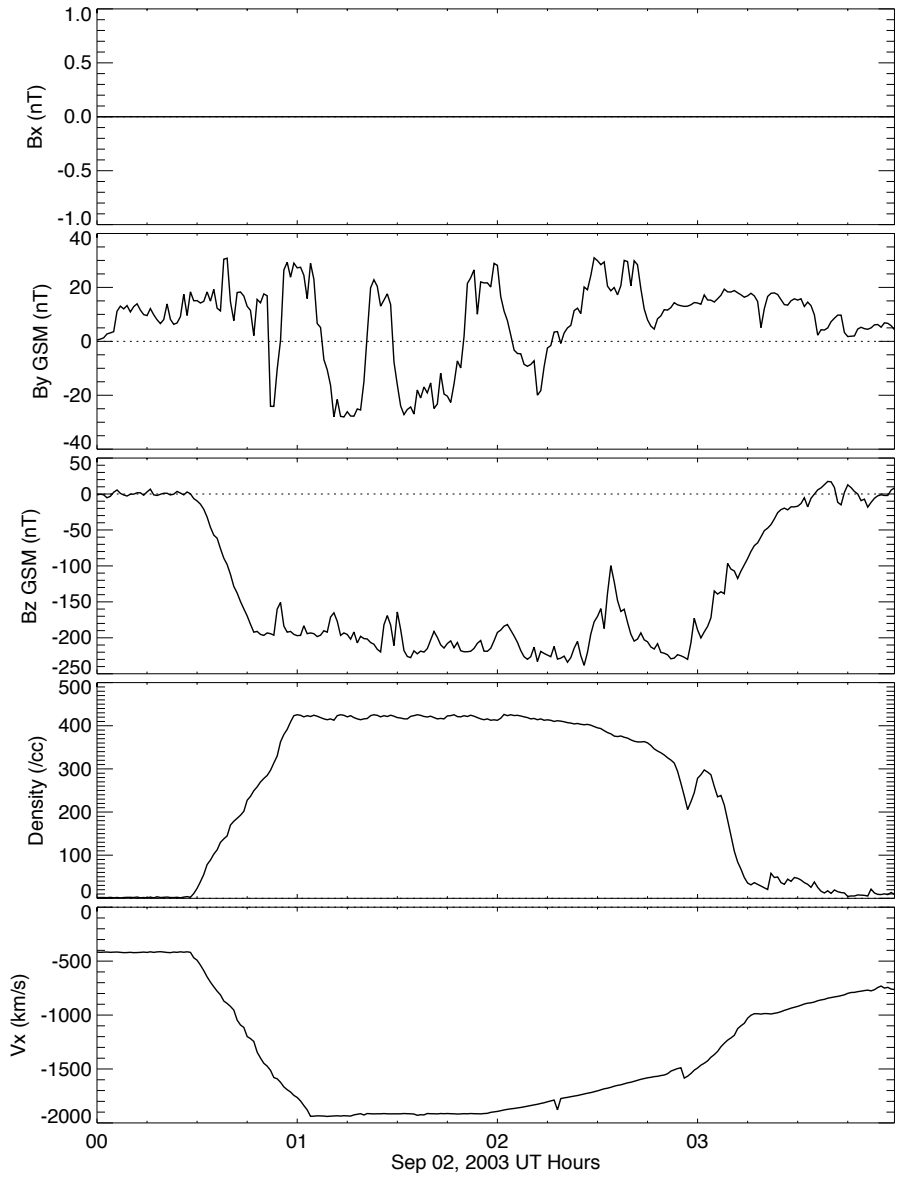}
\caption{Time evolution of the three components of solar wind/CME's magnetic field $B_x$, $B_y$ and $B_z$, the plasma density and $V_x$ velocity component}.
\end{figure}

Figure 2 presents a 2D map of the steady-state plasma density superimposed by magnetic field lines for the magnetospheric configuration in the Y=0 plane corresponding to the initial 30 minutes of the simulations, when the Earth's magnetosphere was driven only by dynamic pressure from the solar wind. The figure show the formation of the bow shock at the stand-off distance of ~12~$R_E$, which is typical for the solar wind conditions.

\begin{figure}
\centering
\includegraphics*[width=0.5\linewidth, angle=90]{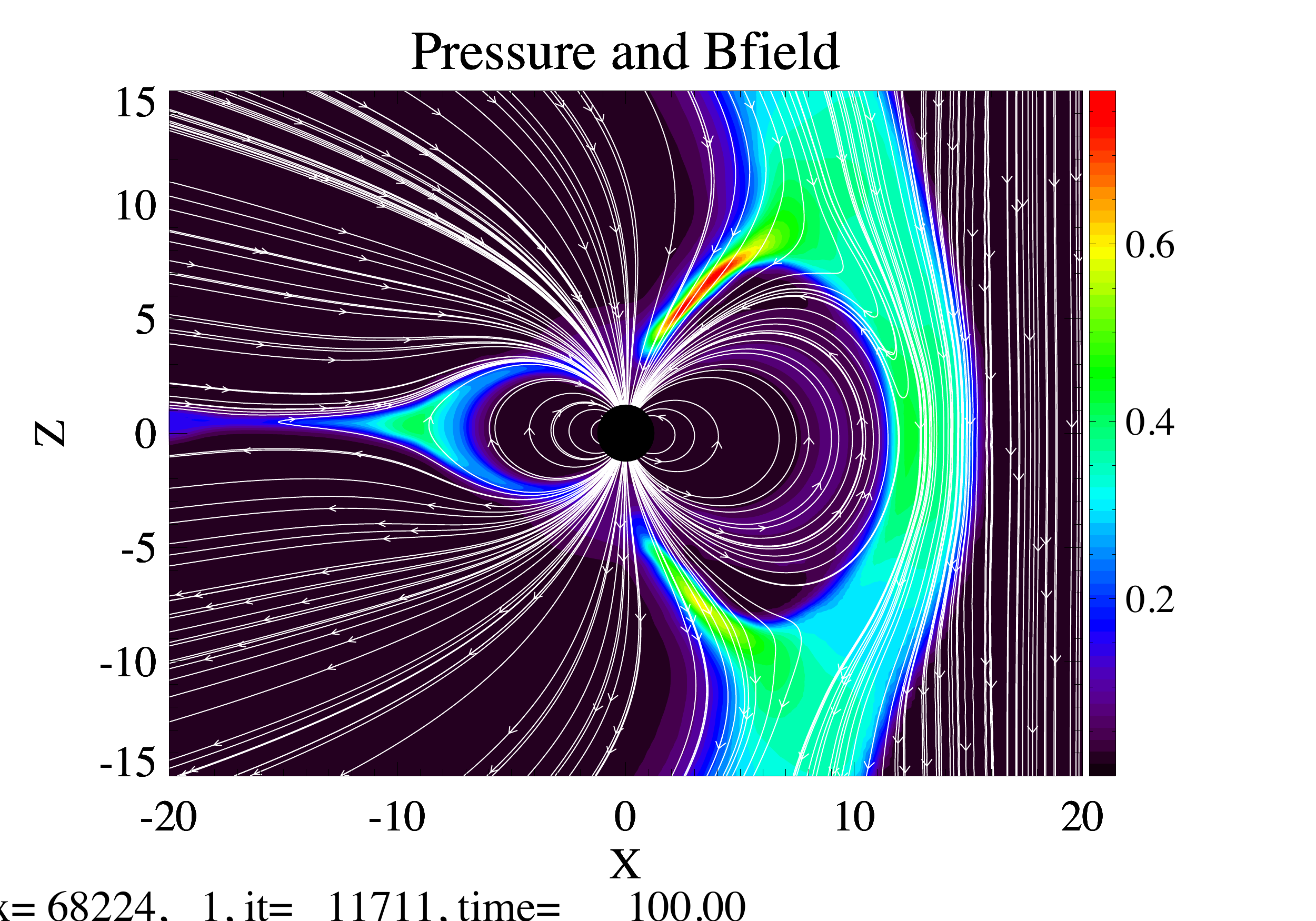}
\caption{Initial conditions.  Plasma pressure and the magnetic field lines.}
\end{figure}

\begin{figure}
\centering
\includegraphics*[width=\linewidth]{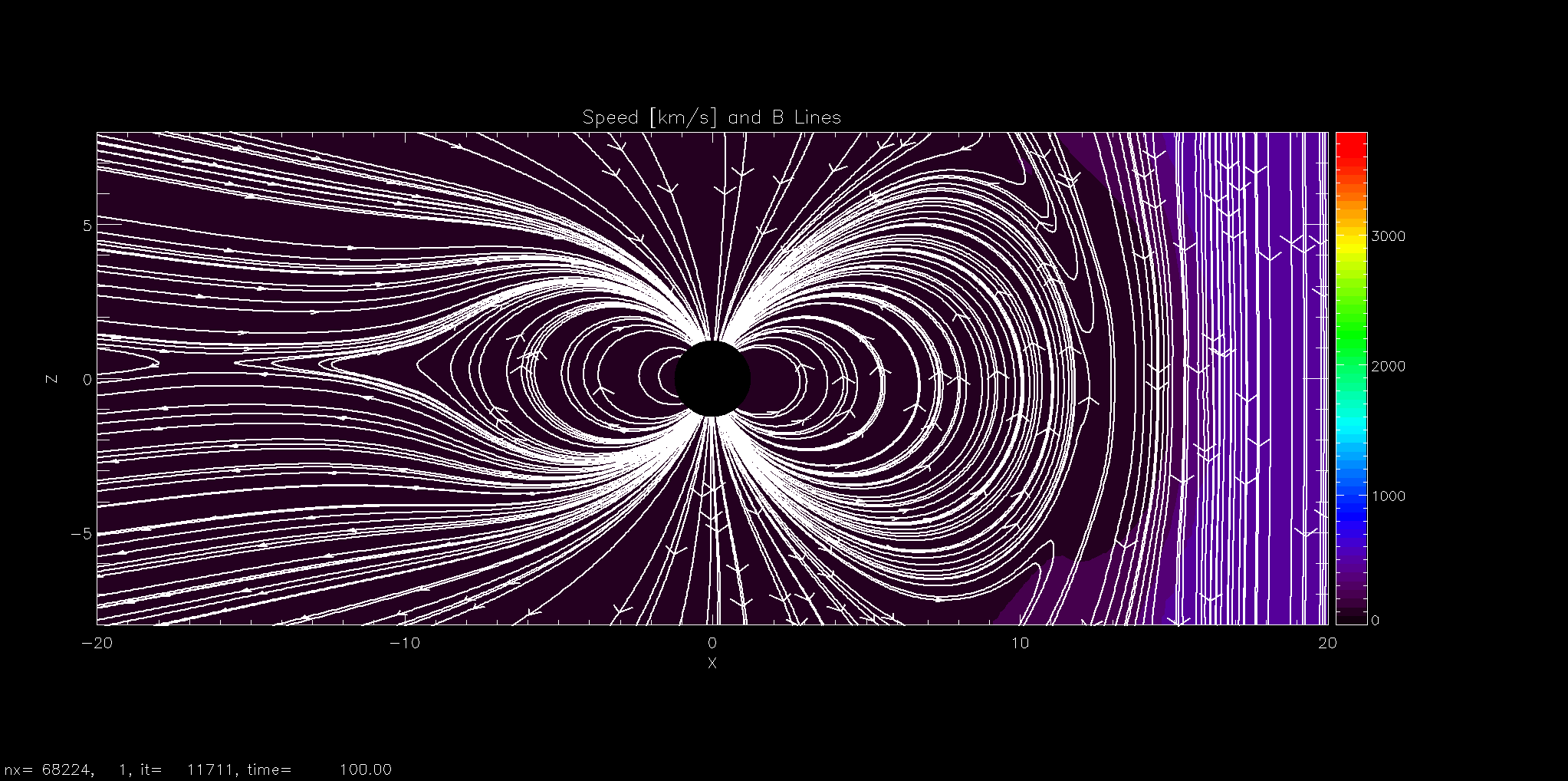}
\caption{Initial conditions.  Plasma velocity and the magnetic field lines.}
\end{figure}

\begin{figure}
\centering
\includegraphics*[width=\linewidth]{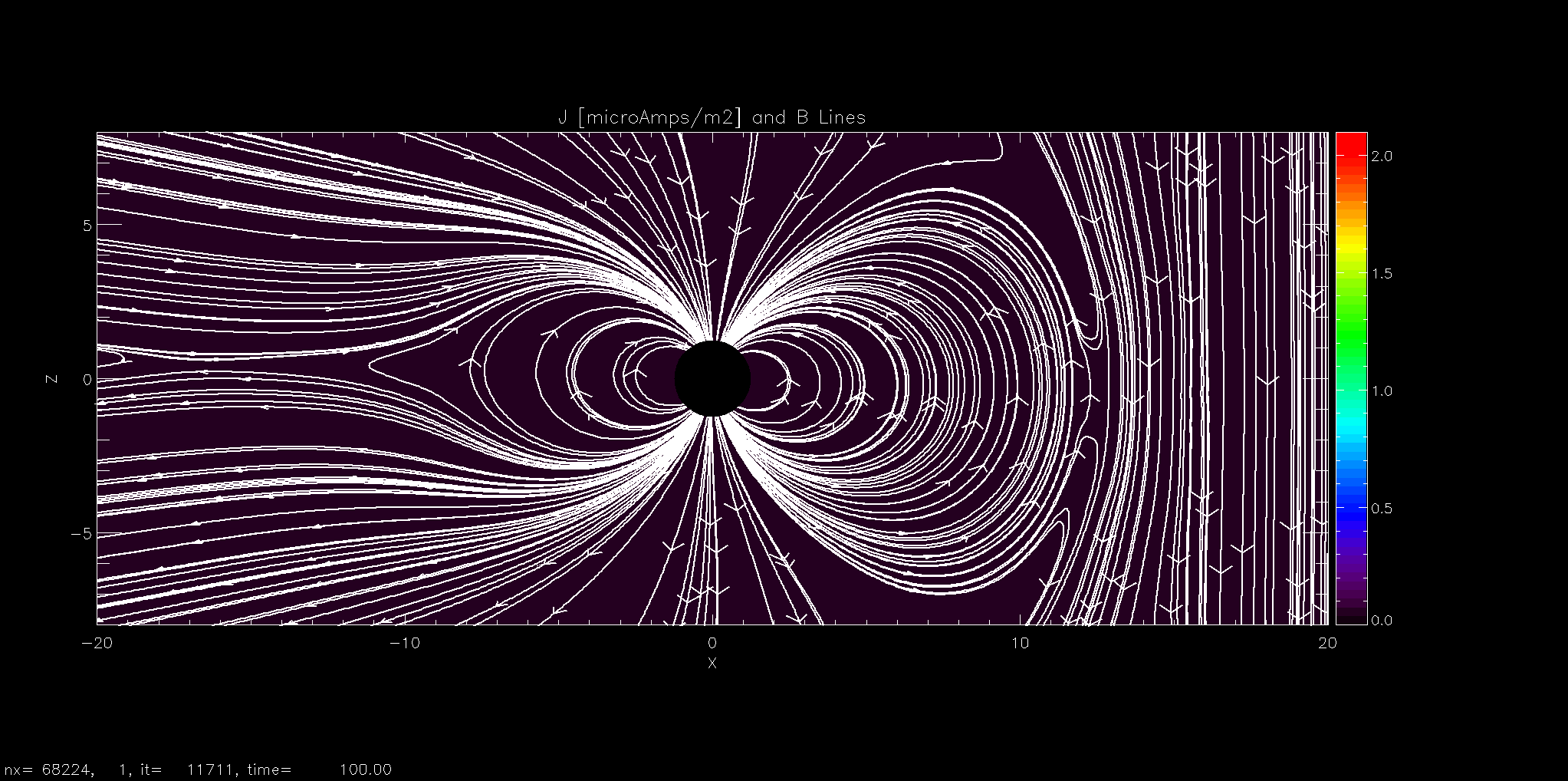}
\caption{Initial conditions.  Current density and the magnetic field lines.}
\end{figure}

\section{SCME impact on Earth's magnetosphere}

\subsection{Magnetospheric response}

At t=30 min, we introduce a SCME event described by the time profiles of y- and z-components of the cloud's magnetic field. $B_y$, $B_z$, the plasma density and the x-component of the cloud's velocity, $V_x$,  as the SCME approaches the Earth. The maximum velocity of the event is 1800 km/s. The SCME magnetic field is directed southward or is sheared by 180 degrees with respect to the dipole field with the $B_z$= -212 nT. As the SCME propagates inward (from right to left in Figures 2 and 3-5), its large dynamic pressure compresses and convects the magnetospheric field inducing the convective electric field. It also compresses the night-side magnetosphere and ignites magnetic reconnection at the nigh-side of the Earth's magnetosphere causing the magnetospheric storm as particles penetrate the polar regions of Earth. Another effects appears to be crucial in our simulations. The strong sheared magnetic field on the day side (sub-solar point) of Earth is also subject to reconnection, which dissipates the outer regions of the Earth's dipole field up until 1.5 $R_E$ above the surface.

Figures 3, 4 and 5 show the snapshots of the density, velocity and current density respectively for the magnetospheric configuration at t=10 hours after the beginning of the magnetic storm. We can see that the density in the magnetosheath increases by factor of about 250, magnetic field - by about 120, and pressure - by about a factor of 4000.

\begin{figure}
\centering
\includegraphics*[width=0.5\linewidth, angle=90]{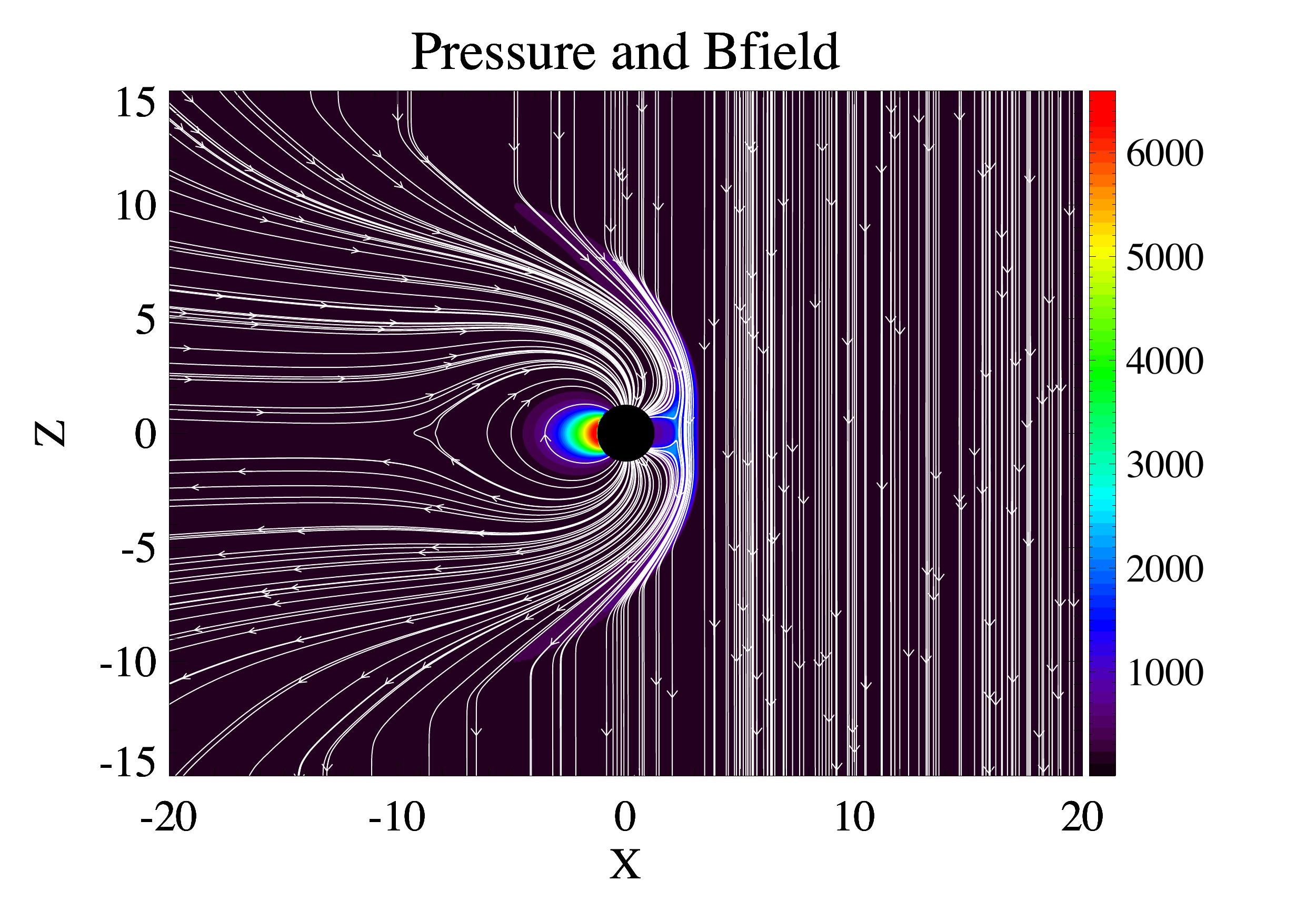}
\caption{Magnetospheric storm at t=10 h. Plasma pressure and the magnetic field lines.}
\end{figure}

\begin{figure}
\centering
\includegraphics*[width=0.5\linewidth]{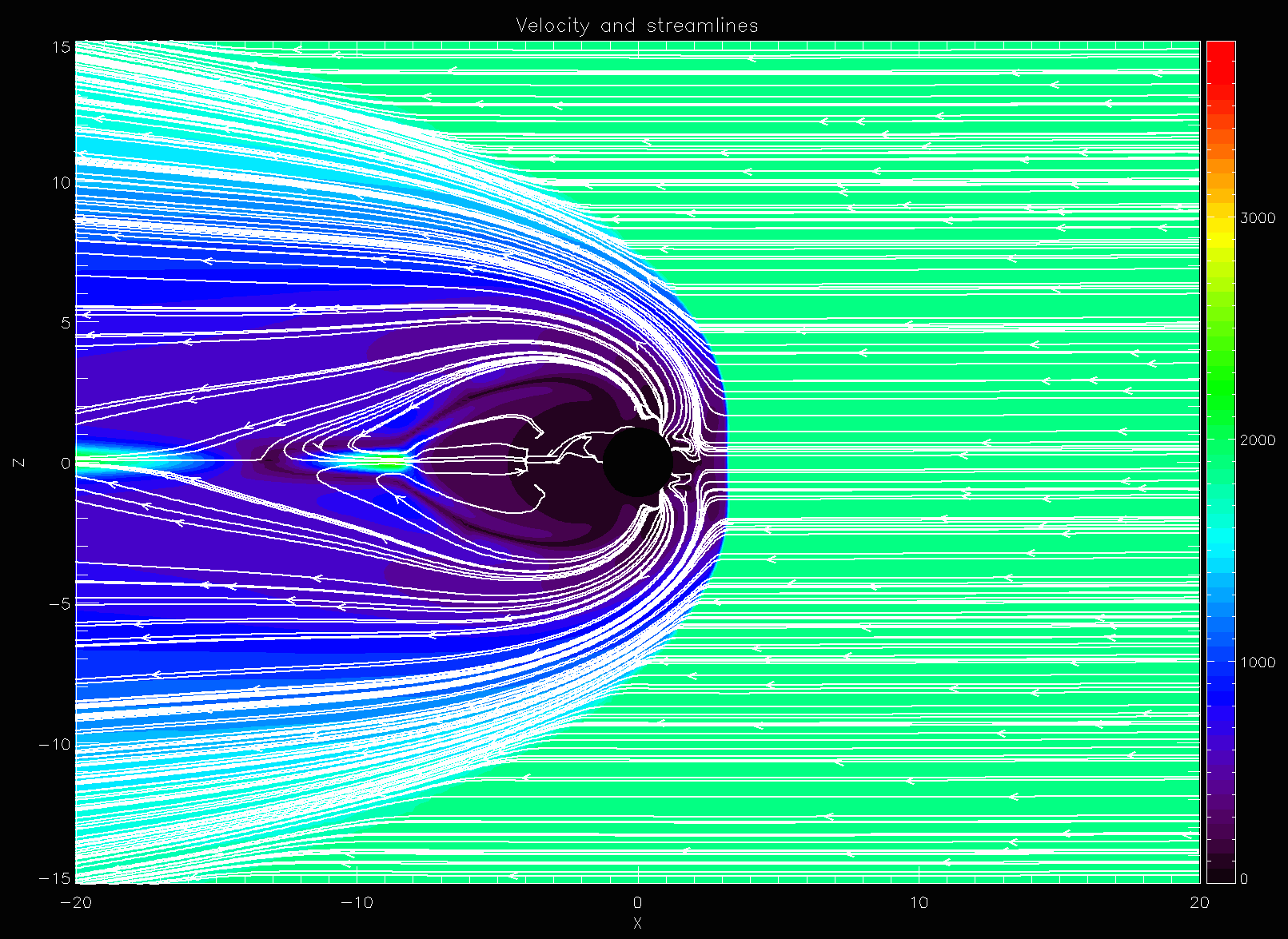}
\caption{Magnetospheric storm at t=10 h. Plasma velocity and the magnetic field lines.}
\end{figure}

\begin{figure}
\centering
\includegraphics*[width=0.5\linewidth]{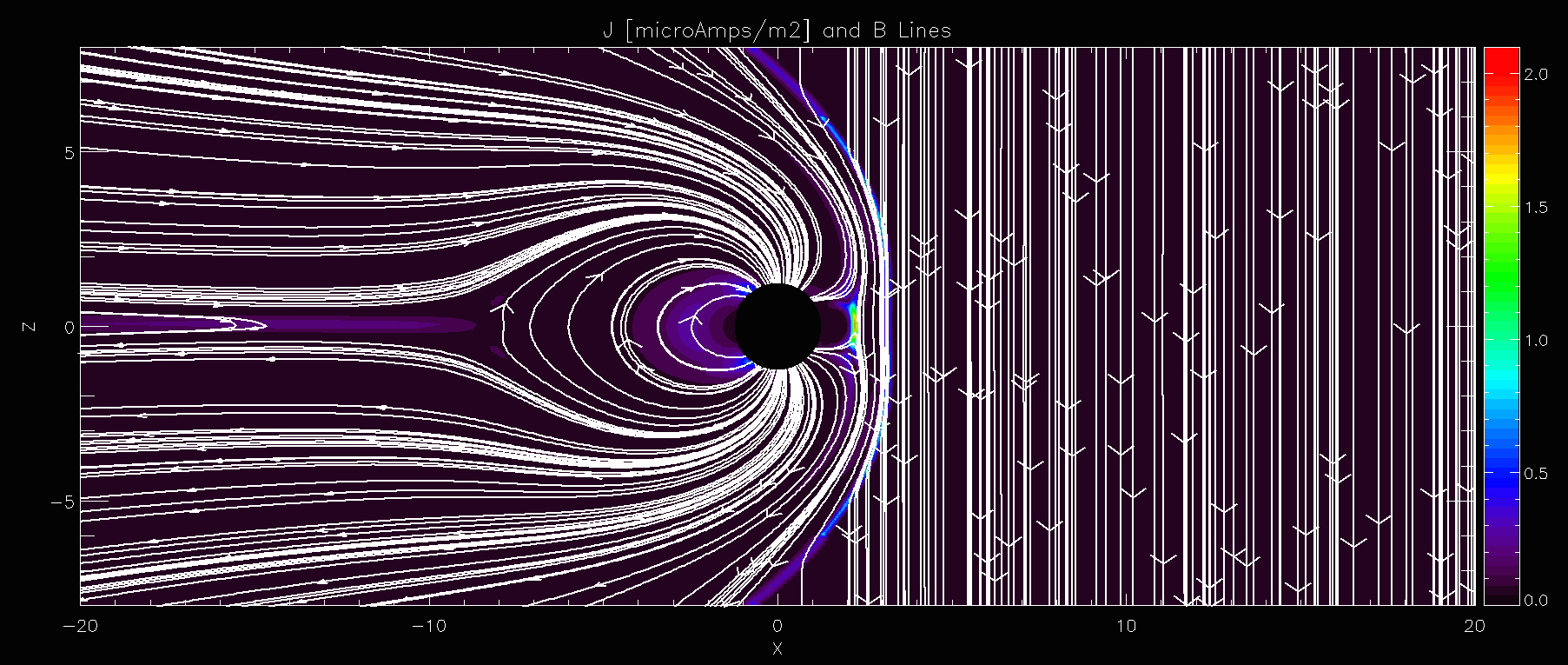}
\caption{Magnetospheric storm at t=10 h. Current density and the magnetic field lines.}
\end{figure}

\subsection{Ionospheric response}

Intensive reconnection of the incoming SCME with the Earth's dipole field at 1 R$_E$ from the surface produces global restructuring of the open magnetic field forms a substantial opening the magnetic field line at the day side that reaches 65$\%$ at 1.3 $R_E$ (see Figure 8). The boundary of the open-closed field shifts to 36 degrees in latitude.

\begin{figure}
\centering
\includegraphics*[width=0.5\linewidth]{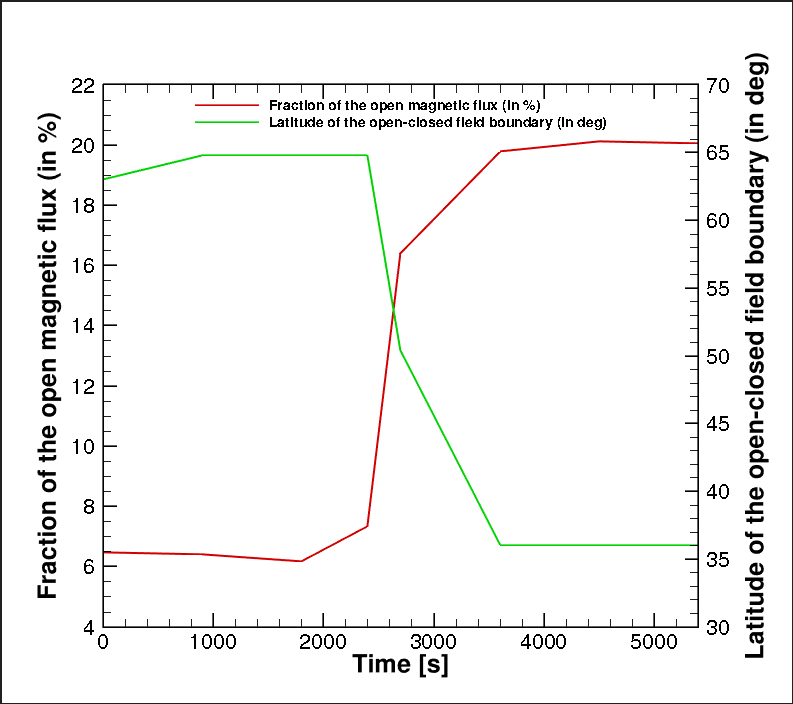}
\caption{The ionospheric boundary at 1.3 $R_E$ at t=10 h: The fraction of the open magnetic flux and the minimum latitude of the open-closed field boundary.}
\end{figure}

The ionospheric response is characterized by the ionospheric potential derived from the field aligned current that was produced by the MHD solution and an ionospheric conductance. The SCME drives large field aligned currents that provide a Joule ionospheric heating reaching 4000 erg/cm$^2$/s as presented in Figure 9. Ionospheric heating large gradients of plasma pressure in addition to {bf $\vec{J}\times \vec{B}$} forces which drive mass outflow at velocities greater than 20 km/s.
Ionospheric cross cap potential drives large energy flux of non-thermal precipitating electrons $\sim$ 24 erg/cm$^2$/s with the mean energy of 5 keV (see Figure 10)

\begin{figure}
\centering
\includegraphics*[width=0.5\linewidth]{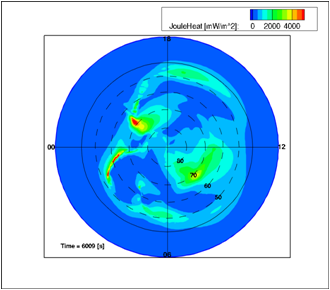}
\caption{Magnetospheric storm at t=10 h. The Joule ionospheric heating rate in W/$m^2$.}
\end{figure}

\begin{figure}
\centering
\includegraphics*[width=0.5\linewidth]{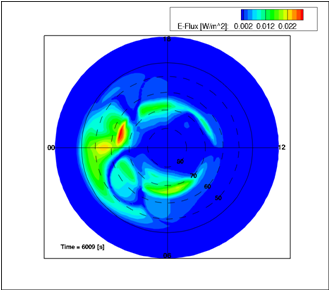}
\caption{Magnetospheric storm at t=10 h. The flux rate of electrons precipitating into the Earth's ionosphere.}
\end{figure}

\section{Implications for Life on Early Earth and Earths Twins}

The simulated CME event is by a factor of 2 greater than the Carrington event, but less energetic than the 775 AD event. Its occurrence rate of 1 event per 800-1000 yr has been estimated from the rate of superflares observed on slow rotating sun-like stars as well as from the statistics of rare events (Love 2012; Shibayama et al. 2013). Such event if accompanied by the negative B$_z$ magnetic field orientation with respect to the Earth's dipole field, would cause catastrophic consequences for our technological society and serious implications for life on Earth. The proton fluence with energies $>$ 30 Mev is estimated to be as high as 10$^{12}$ erg/cm$^2$ using the observed fluences from one of the strongest SEP events on record, the October 1989 SPE (Thomas et al. 2013). Our simulations suggests that due to intensive reconnection, about 65$\%$ of the Earth's dipole magnetic field opens at ~ 1.3 R$_E$ from its surface. This suggests that energetic protons will penetrate the Earth's atmosphere causing atmospheric ionization due to collisions efficiently producing NO$_x$ (N, NO and NO$_2$), which deplete ozone. Simulations by Thomas et al (2013) suggest that this process can increase UVB solar irradiance by as high as 317$\%$ in the case of the soft spectrum of protons accelerated in the CME event. This would produce a serious damage to a human skin (in the form of erythema), disruption in terrestrial plant growth. If we extend these simulations to the scenario of a continuous exposure of young Earth's magnetosphere to SCMEs with energies comparable to the AD 775 event and take into account that the young Sun had produced superflares ($>$ 10$^{33}$ erg) at the occurrence rate of a few events per day, this would correspondingly increase the heating rate of the Earth atmosphere due to radiative heating as well as increase UVB irradiation by two orders of magnitude with respect to the AD 775 event. We expect that such high radiative output from the young Sun and the proton fluence from SCMEs continuously hitting the Earth's magnetosphere would have detrimental effects for life on early Earth in the first billion years after its formation. We are planning to expand our studies to the case of the early Sun - Earth magnetic interaction with corresponding assessment of effects on the early Earth atmosphere and conditions on its living forms.

%
% NTA: Information here on citations using \citep and \citet
%

%
% NTA: Here is an example JPG illustration.  This only works under PDFLaTeX, not straight LaTeX.
%

%\begin{figure}[ht!]
%\centering
%\includegraphics[width=90mm]{cs18-bags-photo.JPG}
%\caption{Caption for an example JPG image.  Lot of CS18 bags!}
%\label{overflow}
%\end{figure}

%\acknowledgments{
%Any acknowledgements you see fit.
%}

\normalsize

%\begin{references}{}
{}

\end{document}